\documentclass[twocolumn,preprintnumbers,amsmath,amssymb,prl]{revtex4}

\usepackage{xspace}
\usepackage{graphicx}% Include figure files
\usepackage{dcolumn}% Align table columns on decimal point
\usepackage{bm}% bold math
\usepackage{tabularx}% Table width
\usepackage{amsmath}% Higher math
\usepackage{color}%color
\usepackage{ulem}

\begin{document}

\title{Spinon heat transport in the three-dimensional quantum magnet PbCuTe$_2$O$_6$}

\author{Xiaochen Hong$^{1,2}$, Matthias Gillig$^{2}$, Abanoub R. N. Hanna$^{3,4}$, Shravani Chillal$^{4}$, A. T. M. Nazmul Islam$^{4}$, Bella Lake$^{3,4}$, Bernd B\"uchner$^{2,5}$, Christian Hess$^{1,2,\footnote{c.hess@uni-wuppertal.de}}$}

\affiliation{
$^1$Fakult\"at f\"ur Mathematik und Naturwissenschaften, Bergische Universit\"at Wuppertal, Gau{\ss}stra{\ss}e 20, 42119 Wuppertal, Germany\\
$^2$Leibniz-Institute for Solid State and Materials Research (IFW-Dresden),  Helmholtzstra{\ss}e 20, 01069 Dresden, Germany\\
$^3$Institut f\"ur Festk\"orperforschung, Technische Universit\"at Berlin, Hardenbergstr. 36, 10623 Berlin, Germany\\
$^4$Helmholtz-Zentrum Berlin f\"ur Materialien und Energie, Hahn-Meitner Platz 1, 14109 Berlin, Germany\\
$^5$Institute of Solid State and Materials Physics and W\"urzburg-Dresden Cluster of Excellence $ct.qmat$, Technische Universit\"at Dresden, 01062 Dresden, Germany\\
}

\date{\today}

\begin{abstract}
Quantum spin liquids (QSL) are novel phases of matter which remain quantum disordered even at the lowest temperature.
They are characterized by emergent gauge fields and fractionalized quasiparticles.
Here we show that the sub-Kelvin thermal transport of the three-dimensional $S=1/2$ hyper-hyperkagome quantum magnet PbCuTe$_2$O$_6$ is governed by a sizeable charge-neutral fermionic contribution which is compatible with the itinerant fractionalized excitations of a spinon Fermi surface.
We demonstrate that this hallmark feature of the QSL state is remarkably robust against sample crystallinity, large magnetic field, and field-induced magnetic order, ruling out the imitation of QSL features by extrinsic effects.
Our findings thus reveal the characteristic low-energy features of PbCuTe$_2$O$_6$ which qualify this compound as a true QSL material.
\end{abstract}

\pacs{not needed}

\maketitle
Quantum spin liquid (QSL) states refer to highly entangled magnetic quantum ground states realized in frustrated magnets \cite{Balents2010,Savary2016}.
Despite the quantum disorder of the ground states, the QSL possess well defined emergent fractionalized excitations such as spinons, Majorana fermions, visons, and many more  \cite{Balents2010,Savary2016,Zhou2017,Broholm2020,Wen2002,Wen2019}, rendering them tantalizing since their initial proposal \cite{Anderson1973}.
Recent years have witnessed the progress of materializing the QSL models \cite{Lee2008}, which has stimulated intense interest from both experimental and theoretical sides \cite{Zhou2017,Broholm2020}.
By far most efforts in this field are devoted to two-dimensional systems because enhanced quantum fluctuations, an ingredient for realizing QSL states, are prominent in reduced dimensionality \cite{Savary2016,Zhou2017,Broholm2020}.
Nevertheless, there are also some three-dimensional (3D) QSL candidates. The most prominent model systems are examples of pyrochlore, hyperkagome, and double-layer kagome lattices \cite{Gingras2014,Okamoto2007,Gao2019,Balz2016,Plumb2018}.

Among the handful of candidate QSL materials, clear-cut evidence for the anticipated emergent fractionalized magnetic excitations, in particular the spinons, is rather scarce \cite{Zhou2017,Broholm2020}.
Thermal conductivity, a probe only sensitive to itinerant entropy carriers, is the method of choice to prove the existence of spinons via their fermionic nature and their mobility \cite{Lee2007, Yamashita2012}.
These important pieces of information are difficult to diagnose by thermodynamic or spectroscopic studies.
To be specific, the spinon contribution to the heat conductivity $\kappa_\mathrm{spinon}$ is expected to be linear in temperature ($T$) towards $T\rightarrow 0$~K \cite{Yamashita2012}, reminiscent of the electronic $\kappa_\mathrm{e}$ in metals.
Except some one-dimensional spin-chain systems \cite{Hess,Pan,Hlubek}, compelling experimental evidence for this sought-after $\kappa_\mathrm{spinon}$ signalling a spinon Fermi surface is still pending.
Earlier reports for $\kappa_\mathrm{spinon}$ in other QSL candidate materials have been shown to suffer from irreproducibility \cite{Yamashita2010,Yu2017,Murayama2020,Ni2019,BourgeoisHope2019}. Some other QSL-like results can actually be explained by a peculiar phononic background resulting from spin scattering \cite{Li2020,Rao2021,Guang,NCTO}.
It is also argued that defects and impurities in a genuine QSL material can easily eliminate its fingerprints in thermal transport \cite{Murayama2020,Yamashita2020a,Huang2021}, preventing a reconciliation with other experimental techniques with which disorders do just the opposite, producing fraudulent QSL-like results in trivial systems \cite{Zhu2017,Chamorro2021,Ma2020}.
It is therefore essential to exclude all these problems in order to reveal a true evidence for fermionic spinon heat transport, $\kappa_\mathrm{spinon}$.

Choloalite PbCuTe$_2$O$_6$ crystallizes into a cubic P4$_1$32 structure at ambient temperature \cite{Inosov2018,Koteswararao2014}.
Its magnetic Cu$^{2+}$ ions ($S=1/2$) constitute a 3D network, similar to the Ir$^{4+}$ ions in the hyperkagome material Na$_4$Ir$_3$O$_8$ \cite{Okamoto2007}.
However, density function theory calculations of PbCuTe$_2$O$_6$ suggest its nearest neighbor ($J_1 = 1.13$~meV, giving isolated triangles) and next nearest neighbor ($J_2 = 1.07$~meV, giving a hyperkagome lattice) interactions are almost of the same strength \cite{Chillal2020}.
As a result, each Cu$^{2+}$ ion is at the corner of three triangles rather than two triangles as for the hyperkagome case. The Cu$^{2+}$ lattice of PbCuTe$_2$O$_6$ thus possesses 4-site and 6-site loops as its shortest spin rings \cite{Chillal2020}, distinct from the 10-site loop of a standard hyperkagome lattice \cite{Zhou2008,Bergholtz2010}, and was referred to as hyper-hyperkagome lattice \cite{Chillal2020}.
An antiferromagnetic Curie-Weiss temperature $\Theta_\mathrm{CW}\approx -22$~K was inferred from the magnetic susceptibility data of PbCuTe$_2$O$_6$ \cite{Koteswararao2014}, but no magnetic order has been found down to 20~mK (frustration parameter $f=|\Theta_\mathrm{CW}/T_N|>1000$, where $T_N$ is the N\'{e}el temperature) \cite{Khuntia2016,Koteswararao2014,Chillal2020}.
PbCuTe$_2$O$_6$ possesses a ferroelectric (FE) transition at around 1~K, accompanied by a structural transition to a lower symmetry phase \cite{Thurn2021,Hanna2021}.
Notably, both the FE and structural distortions are absent in small-grained polycrystalline samples \cite{Hanna2021}.
Regardless of this difference, characteristic multi-spinon continua of the magnetic excitations were identified by inelastic neutron scattering in both polycrystalline and single crystalline PbCuTe$_2$O$_6$ samples \cite{Chillal2020}, down to 100~mK (below $T_\mathrm{FE}$). This places PbCuTe$_2$O$_6$ on the shortlist of promising QSL materials with emergent spinon excitations.

\begin{figure}[h]
\centerline{\includegraphics[width=0.44\textwidth]{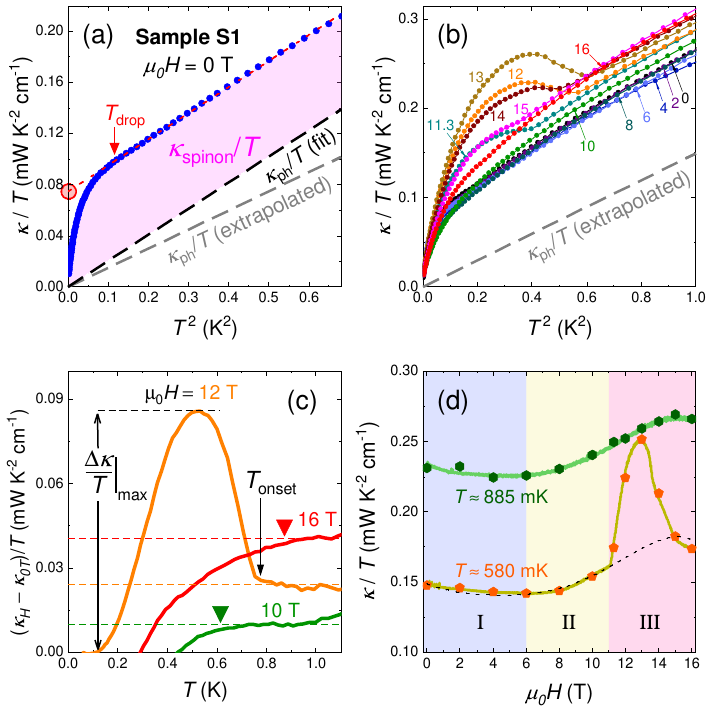}}
\caption{%$\kappa/T$ of Sample S1.
(a) $\kappa/T$ versus $T^2$ of Sample S1 in zero field, and extraction of $\kappa_\mathrm{ph}$ and $\kappa_\mathrm{spinon}$, see text.
The red dotted line represents a linear fit to $\kappa/T$ above $T_\mathrm{drop}$ (red arrow), and its extrapolation to $T=0$, yielding the linear residual  $a\equiv\kappa_\mathrm{spinon}/T$ (red dot).
The dashed black line represents the thus extracted $\kappa_\mathrm{ph}/T$. The spinon contribution $\kappa_\mathrm{spinon}/T$ on top of $\kappa_\mathrm{ph}/T$ is highlighted as the magenta region. 
%fitted linearly above $T_\mathrm{drop}$ (red arrow) as represented by the dotted red line.
%a linear fit to the data above $T_\mathrm{drop}$, indicated by the red arrow.
%(a) The temperature dependence of thermal conductivity below 0.8~K in zero field (blue bullets), plotted as $\kappa/T$ versus $T^2$. The dotted red line represents a linear fit to the data above $T_\mathrm{drop}^2=0.12~\mathrm{K}^2$.
%$T_\mathrm{drop}\approx 340$~mK is indicated by the red arrow.
%The dashed black line stands for phononic contribution $\kappa_{ph}/T$, see text, extracted from the fit.
The dashed gray line represents the extrapolated $\kappa_{ph}/T$ from the $\kappa(T)$ data above 6~K \cite{SM,Note}.
%The spinon contribution $\kappa_\mathrm{spinon}/T$ on top of $\kappa_{ph}/T$ is underlined as the magenta region.
(b) $\kappa/T(T^2)$ in field together with the same extrapolated $\kappa_{ph}/T$ curve as in panel (a).
(c) Field effect on $\kappa$ for three representative field values after subtracting the zero field value $(\kappa_H-\kappa_{0T})/T$.
$T_\mathrm{drop}$ is highlighted by triangles, and the definition of the peak height ($\frac{\Delta\kappa}{T}|_\mathrm{max}$) is exemplified.
(d) Field dependence of $\kappa/T$ isotherms measured at two selected temperatures (thick solid lines), plotted with the results (full symbols) extracted from the fixed-field $\kappa(T)/T$ data. Three different regions can be identified and are displayed by the different background colours. The black dashed line represents the higher temperature isotherm (green band) subtracted by a fixed value of $0.085~\mathrm{mW}/(\mathrm{K}^2\mathrm{cm})$, see text.
}
\end{figure}

In this Letter, we report clear evidences of spinon heat transport in this 3D QSL candidate material, PbCuTe$_2$O$_6$, at very low temperature, revealed by a sizeable $T$-linear contribution ($\kappa_\mathrm{spinon}$) to the total thermal conductivity $\kappa$, which adds to the well-known $T^3$ contribution of phonons ($\kappa_\mathrm{ph}$) \cite{Thacher1967}.
Three different batches of PbCuTe$_2$O$_6$ samples prepared by different techniques, namely two differently fabricated single crystals and one polycrystalline sample were involved in this study \cite{SM}.
$\kappa$ of all samples has been studied at $T<1$~K as well as at $6~\mathrm{K}\leq T \leq 160~\mathrm{K}$ \cite{SM}.
In spite of a rich $H-T$ phase diagram of PbCuTe$_2$O$_6$, the signature of spinon heat transport is robust in all samples across the investigated parameter range.

\begin{figure*}
\centerline{\includegraphics[width=0.7\textwidth]{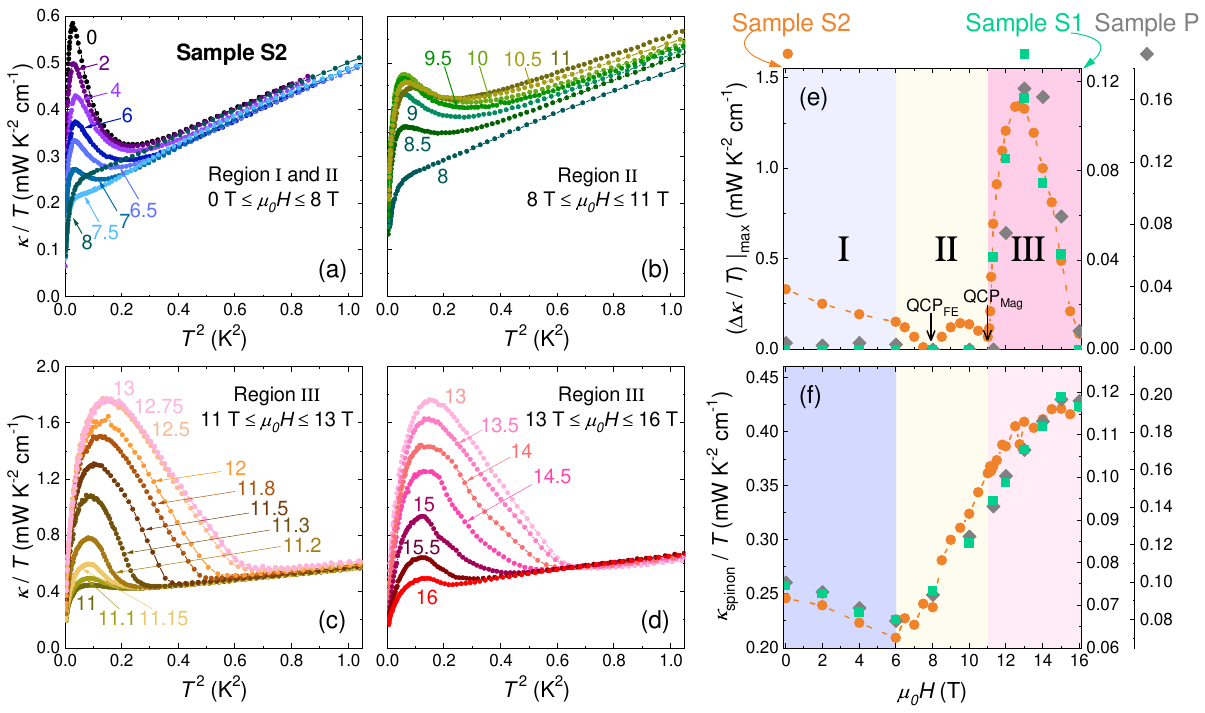}}
\caption{The $\kappa/T$($T^2$) curves of Sample S2, separated into four panels (a-d) for clarity.
%and the comparison with Sample S1.
%For clarity, the $\kappa/T$($T^2$) curves up to 16~T are separated into four panels (a-d).
%The thermal conductivity of Sample S2 in magnetic field up to 16~T is plotted in $\kappa/T$ as function of $T^2$. For clarity, the curves are separated into four panels, (a) $0 \leq \mu_0H \leq 8~\mathrm{T}$, (b) $8~\mathrm{T}\leq \mu_0H \leq 11~\mathrm{T}$, (c) $11~\mathrm{T}\leq \mu_0H \leq 13~\mathrm{T}$, and (d) $13~\mathrm{T}\leq \mu_0H \leq 16~\mathrm{T}$.
The field dependence of the extracted amplitude of (e) the peaks and (f) the residual linear term $\kappa_0/T$ are compared between all samples.
The ferroelectric (QCP$_\mathrm{FE}$) and magnetic (QCP$_\mathrm{Mag}$) critical points \cite{Eibisch2023} are marked by the black arrows.
Note for the distinct ordinates with different scales.
%for Sample S2 (left) and Sample S1 (right) respectively. They differ by a factor of about 12 in (e) and 3.4 in (f). 
The determination of field regions is inherited from Fig.~1(d). }
\end{figure*}

As presented in Fig.~1(a), the low-$T$ $\kappa(T)$ of the single crystalline Sample S1 at zero field shows distinct fingerprints of a spinon Fermi surface, $i.e.$ a linear contribution to $\kappa$.
This can easily be seen from the figure which shows the measured $\kappa/T$ versus $T^2$.
In this representation, the linear contribution on top of a standard phononic background ($\kappa_{ph}\propto T^3$ \cite{Berman}) is just the residual at 0 K \cite{Yamashita2012}.
Indeed, above a certain $T_\mathrm{drop}\approx340$~mK, the data can be well fitted according to $\kappa/T=a+b \times T^2$.
Here $bT^2 \equiv \kappa_{ph}/T$ yields the expected phononic background with $b=0.206~\mathrm{mW}/(\mathrm{K^4cm}$) \cite{Note}.
The residual $a=0.075~\mathrm{mW}/(\mathrm{K^2cm})$, as indicated by the red dot, represents a fermionic contribution to the heat conductivity.
For insulating magnets such as PbCuTe$_2$O$_6$, such a fermionic contribution can only be explained by pertinent fractionalized magnetic excitations \cite{Yamashita2010,Lee2007}.
The data thus provide clear-cut evidence for a spinon contribution, $\kappa_\mathrm{spinon} \equiv aT$.
%$\kappa_\mathrm{spinon}$ is sizeable and even exceeds $\kappa_{ph}$ as is highlighted by the magenta shaded area.
%Thus, for $T>T_\mathrm{drop}$, our data evidence a spinon Fermi surface in PbCuTe$_2$O$_6$.
%The origin of the decay of $\kappa_\mathrm{spinon}$ at $T<T_\mathrm{drop}$ will be discussed at the end of the paper.

Our above conclusion of spinon heat transport and its quantitative determination is corroborated by our measurements of $\kappa/T$ in magnetic fields.
Fig.~1(b) shows $\kappa/T(T)$ of Sample S1 in magnetic fields up to $\mu_0H=16$~T, with the field applied perpendicular to the heat current.
The curves are only slightly affected by the field up to $\mu_0H=10$~T, and are shifted up with a nearly unaltered $bT^2$ term at higher fields.
This renders our data fundamentally different from recently claimed evidences for a spinon residual $\kappa_\mathrm{spinon}/T$ term in several QSL candidate materials \cite{Rao2021,Li2020,Guang}, where a large magnetic field leads to a strong enhancement (factor $2 ... 10$) of $\kappa$ and $bT^2$ \cite{Gillig2021,Hong2022,NCTO}.
%, evidencing a strong phonon-spin scattering \cite{Gillig2021,Hong2022,NCTO}, which casts doubts on a spinon transport origin of the residual $a$ in zero field in these cases.
Contrastingly, the practically field independent $bT^2$ term in PbCuTe$_2$O$_6$ and the fact that the pure $\kappa_{ph}/T(T)$ curve is well below the total $\kappa/T(T)$ curves leave no room for a phononic-only explanation to the residual $a$ as found in other frustruated magnets \cite{NCTO}, confirming the $\kappa_\mathrm{spinon}$ transport channel.

We point out that while $\kappa_\mathrm{ph}$ remains weakly affected by the magnetic field, it still has a clear impact on the total $\kappa$.
Firstly, as represented by the 12~T curve in Fig.~1(c) and also visible directly in Fig.~1(b), an additional $\kappa/T(T)$ peak on top of the phonon and spinon contributions emerges below $T_\mathrm{onset}$ in the field range of $11$~T$<\mu_0H<15$~T.
Both the $T_\mathrm{onset}$ and the amplitude of the new peak are highly field sensitive.
Eventually, at $\mu_0H=16$~T the peak is absent again and the low field behavior of $\kappa/T(T)$ is recovered with a rather higher value of $T_\mathrm{drop} \approx$~870~mK.
Secondly, $T_\mathrm{drop}$, which apparently represents an energy scale above which $\kappa_\mathrm{spinon}$ can be observed, increases with the field. As an example shown in Fig.~1(c), $T_\mathrm{drop} \approx$~610~mK at $\mu_0H=10$~T .
Finally, the residual $a\equiv\kappa_\mathrm{spinon}/T$, increases considerably in this high field region above 10~T.

The general field evolution of $\kappa/T$ is more clearly presented in Fig.~1(d) as $\kappa/T(H)$ isotherms.
The high-$T$ (885 mK, above $T_\mathrm{onset}$) isotherm shows a minor decrease at small fields, followed by a slight continuous increase of $\kappa/T$ above $\mu_0H\approx 6~T$, until its saturation around $\mu_0H=15$~T.
This increment is entirely due to $\kappa_\mathrm{spinon}$, as was discussed above.
On the other hand, the lower-$T$ (580 mK, below $T_\mathrm{onset}$) isotherm is more complicated, featuring a large hump centered around 13~T.
Three field regions can thus be discerned based on the isotherms.
The $\kappa/T(H)$ isotherms first decrease mildly in Region I, and then increase somewhat faster in Region II.
Region III is defined by the occurrence of the peak.
It is worthwhile to point out that the two isotherms match nicely in Region I \& II through shifting by a constant value of $0.085~\mathrm{mW}/(\mathrm{K^2cm})$, i.e., the difference in the phononic contribution $b\times[(0.885$~K$)^2-(0.58$~K$)^2]$.
This fact strongly suggests a new contributor to $\kappa$ comes into play below $T_\mathrm{onset}$ in Region III, which will be discussed further below.

In order to evaluate whether any of the observed heat transport phenomenology described above is affected by the ferroelectric and accompanying structural transitions at $T_\mathrm{FE}\approx$1~K, we performed another heat transport measurement on a sample lacking the FE transition, namely the unannealed polycrystalline Sample P.
With regard to the heat transport, it behaves basically the same as the single crystalline Sample S1 in the same $T$ and $H$ parameter range \cite{SM}.
Thus, our above observations represent the intrinsic phenomenology of PbCuTe$_2$O$_6$, which is apparently independent of the ferroelectric or structural transitions.
Note, that our $\kappa/T(T)$ curves show no anomaly at around 1~K \cite{SM}, in contrast to the specific heat, which underpins this statement \cite{Hanna2021,Thurn2021,Eibisch2023,Note2}.
This finding is important, because it demonstrates that the inferred $\kappa_\mathrm{spinon}$ is unaffected by the symmetry reduction induced by the ferroelectric order and the accompanying structural phase transition.
Hence, our conclusion of compatibility with a QSL ground state remains robust even if the subtle non-cubic distortion present in the single crystalline Sample S1 below 1~K is taken into account \cite{Thurn2021}.

After having established the intrinsic low-temperature heat transport behaviour of PbCuTe$_2$O$_6$, we turn now to the single crystalline Sample S2 which (unlike the phase-pure samples Sample S1 and Sample P \cite {SM}) is known to contain small amount of non-magnetic Pb$_2$Te$_3$O$_8$ inclusions in an otherwise phase pure PbCuTe$_2$O$_6$ matrix \cite{Chillal2020,Hanna2021}.
The most obvious difference of the $\kappa(T)/T$ curves of Sample S2, shown in  Fig.~2(a-d), compared to Samples S1 and P is an additional $\kappa/T(T)$ peak below $\mu_0H = $11~T (in Region I\&II).
%The most obvious difference compared to Sample S1 is an additional $\kappa/T(T)$ peak below $\mu_0H = $11~T (in Region I\&II).
%The onset temperature of this additional peak is quite low ($T_\mathrm{onset} \approx$~600~mK), and is little affected by the field.
%On the other hand, the peak amplitude is strongly suppressed with increasing field below $\mu_0H \approx 8$~T, followed by a mild recovery in the range of 8~T~$< \mu_0H < $10~T and then a decrease above 10~T.
It is highly sensitive to $H$, as embodied more clearly in Fig.~2(e), where the extracted peak height of Sample S2 reveals a sharp dip at $\mu_0H \approx$~8~T, exactly the field at which the ferroelectric transition is driven to its critical point ($H_{QCP(FE)}=7.9$~T)\cite{Eibisch2023}. On the other hand, it is insensitive to the boundary between Region I and Region II.
At higher field, the additional peak is preempted by the feature bounded to Region III.
Although the $\kappa/T(T)$ peak of Sample S2 in Region III is much more pronounced as compared to Sample S1 and P, their extracted contributions ($\Delta\kappa/T|_{max}$) match extremely well after normalizing the peaks by their maximum value at 13~T (see Fig.~2(e)).
The field dependent spinon contribution $\kappa_\mathrm{spinon}/T$ is also extracted from the data set above $T_\mathrm{onset}$, and compared between all three samples, as presented in Fig.~2(f).
Again they fit very well modulo a proper rescaling factor.
In all samples, the $\kappa_\mathrm{spinon}/T$ values at 16~T increase by 70\% or more with respect to their 0~T values.
%The observed field dependence of $\kappa_\mathrm{spinon}$ thus is intrinsic. 
Further theoretical investigations are required to rationalize this field dependence. At present, one may speculate that it results from a field-induced variation of the spinon band width \cite{Forte2013}, or a field-induced shift of the spinon chemical potential \cite {Shen2018}.

\begin{figure}
\centerline{\includegraphics[width=0.43\textwidth]{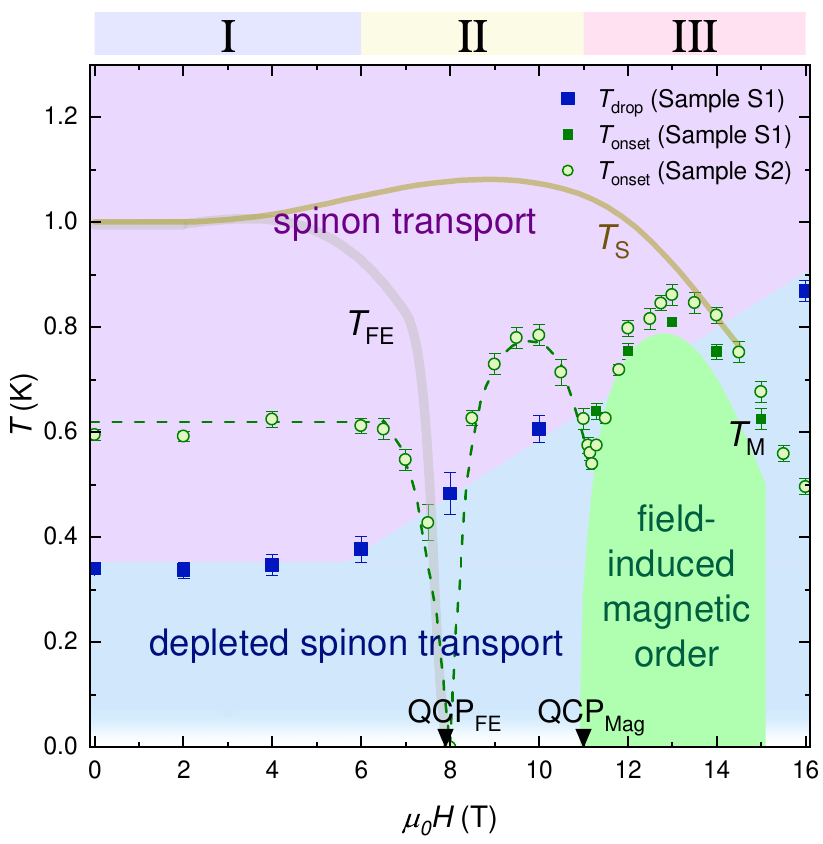}}
\caption{Phase diagram of PbCuTe$_2$O$_6$, showing $T_\mathrm{onset}$ of samples S1 (green squares) and S2 (circles) as well as $T_\mathrm{drop}$ (sample S1) together with $T_\mathrm{FE}$, $T_\mathrm{S}$ (dark yellow and grey lines), and $T_\mathrm{M}$ (top margin of light green area) as extracted from Ref.~\cite{Eibisch2023}. The ferroelectric QCP$_\mathrm{FE}$ at 7.9~T and magnetic QCP$_\mathrm{Mag}$ at 11~T are reproduced from Ref.~\cite{Eibisch2023}. The evidence for a quantum spin liquid state (i. e. finite residual linear term $a$) is not affected by the ferroelectric or structural transitions, thus is ubiquitous in the parameter space coloured in purple. The region below our experimental accessible range at $T\lesssim 50$~mK is left blank, except for the magnetically ordered phase.
}
\end{figure}

It is very revealing to plot our main findings together with a recently established thermodynamic $T-H$ phase diagram of the ferroelectric ($T_\mathrm{FE}$), magnetic ($T_\mathrm{M}$), and structural ($T_\mathrm{S}$) orders at low-$T$ \cite{Eibisch2023}, see Fig.~3.
Clearly, the onsets of the additional $\kappa/T(T)$ peak in Region III ($\mu_0H>$11~T) matches with the thermodynamic magnetic ordering temperatures $T_M$ \cite{Eibisch2023}.
Hence, the excess heat conductivity which causes the peak can unambiguously be attributed to magnon transport.
Note that magnons obey a bosonic behavior and emerge from long-range magnetic order in contrast to the spinons.
At lower fields, the presence of the excess $\kappa/T(T)$ peak is obviously sample dependent since it can only be resolved for Sample S2.
Here, the onset temperature ($T_\mathrm{onset}$) and the peak height (see Fig.~2(e)) track at $H<H_{QCP(FE)}$ the ferroelectric order ($T_\mathrm{FE}$) at somewhat lower $T$ and recover at $H_{QCP(FE)}<H<11$~T in the structurally distorted phase.
It is therefore closely connected with the symmetry reduction due to the ferroelectric order and/or the structural distortion. Since phonons clearly are not sensitive to this symmetry reduction (see above), this low-$T$ peak must be of magnetic origin.
Despite the fact that magnetic order could not be detected below 11~T in previous works \cite{Koteswararao2014, Chillal2020, Khuntia2016, Hanna2021, Thurn2021, Eibisch2023}, we therefore assign also this peak to magnon transport.
It is known that PbCuTe$_2$O$_6$ is proximate to magnetic order \cite{Khuntia2016,Thurn2021}, and our data indicate that the disordered crystal matrix in Sample S2 due to the Pb$_2$Te$_3$O$_8$ inclusions drives this sample to magnetic order at low $T$.
%knowing that PbCuTe$_2$O$_6$ is proximate to a magnetic order \cite{Khuntia2016,Thurn2021}. An extrinsic magnetic order could be driven in the somewhat disordered Sample S2 \cite{SM}.
%This is reasonable, despite the fact that magnetic order could not be detected below 11~T in previous works \cite{Koteswararao2014,Chillal2020,Hanna2021,Thurn2021,Eibisch2023,Khuntia2016}. However, muon spin relaxation and nuclear magnetic resonance data reveal a slowing down of spin fluctuations which imply the proximity of the system to magnetic order \cite{Khuntia2016}. The critical nature is further confirmed by a diverging Gr\"uneisen parameter \cite{Thurn2021}.
%Considering the non-magnetic Pb$_2$Te$_3$O$_8$ second phase inclusions of Sample S2 \cite{Hanna2021}, one may therefore conjecture that weak sample-dependent magnetic order can be induced, if the fine balance between magnetic interactions is disturbed by strain fields around the inclusions of the second phase.
%Apparently, this specialty of Sample S2 together with the symmetry reduction connected with the structural phase transition and the ferroelectric order seems sufficient to tip the system towards magnetic order.
Overall, the most exotic finding of this work, the spinon contribution to thermal transport, prevails throughout the phase diagram until it is either freezing below $T_\mathrm{drop}$, or is overshadowed by a magnetic order.
%We point out that, a linear contribution is not observed in the heat capacity measurement in the sub-Kelvin regime \cite{Khuntia2016,Hanna2021,Thurn2021}, presumably because of the dominance of the entropy associated with the ferroelectric transition (see \cite{SM}). 

Finally, we address the freezing out of $\kappa_\mathrm{spinon}/T$ below $T_\mathrm{drop}$. The depleted $\kappa_\mathrm{spinon}$ signal can either indicate a spinon excitation gap, or the loss of coupling between spinons and the phonon background \cite{Smith}, through which $\kappa$ of an insulator is measured.
Future work is required to clarify which of these two scenarios is valid (see \cite{SM}).

To summarize, the high-quality low-$T$ thermal conductivity of a 3D QSL candidate PbCuTe$_2$O$_6$ strongly suggests the existence of itinerant spinons, and thus of a spinon Fermi surface.
The spinon heat transport is shown to be intrinsic and robust against disorder and field-induced phases.
Our work thus highlights PbCuTe$_2$O$_6$ as an unique model system for QSL research.

%Acknowledgement
We acknowledge fruitful discussions with Wolfram Brenig, Elena Hassinger, and Matthias Vojta. We thank Michael Lang, Bernd Wolf, Paul Eibisch and Christian Thurn for ample discussions about our parallel going experiments and for sharing their thermodynamic results with us before publication.
We further thank Tino Schreiner and Danny Baumann for technical assistance.
This project has been supported by the Deutsche Forschungsgemeinschaft through the Sonderforschungsbereich SFB 1143. Furthermore, this project has received funding from the European Research Council (ERC) under the European Unions' Horizon 2020 research and innovation programme (grant agreement No 647276 -- MARS -- ERC-2014-CoG).


\begin{thebibliography}{99}
\bibitem{Note} Attributing $bT^3$ to $\kappa_\mathrm{ph}$ is endorsed by the very good agreement of the fitted $\kappa_\mathrm{ph}$ and the extrapolated $\kappa_\mathrm{ph}$ (Fig.~1) from the high temperature $\kappa$ data as an independent evaluation, which depends on the dominance of canonical $\kappa_{ph}$ in PbCuTe$_2$O$_6$ at high enough temperatures \cite{SM}.
\end{thebibliography}
\end{document}